\documentstyle[11pt,aaspp4,epsf,rotate]{article}

\newcommand\D{{\cal D}}
\def\psim{\lower.5ex\hbox{$\; \buildrel \propto \over \sim \;$}}
\def\D{{\cal D}}
\def\G{\Gamma}
\def\b{\beta}
\def\to{t_{\rm obs}}
\def\e{\epsilon}
\def\eo{\epsilon_{\rm obs}}

\begin{document}

\title{Short Timescale Variability in the \\ External Shock Model of Gamma Ray Bursts }

\author{Charles D. Dermer\altaffilmark{1} \& Kurt E. Mitman\altaffilmark{2,1}}

\altaffiltext{1}{E. O. Hulburt Center for Space Research, Code 7653,
       Naval Research Laboratory, Washington, DC 20375-5352}
\altaffiltext{2}{Thomas Jefferson High School for Science and Technology, 6560 Braddock
Road, Alexandria, VA 22312}

\begin{abstract}

We have developed a computer model to calculate gamma ray burst (GRB) light curves and
efficiencies from the interaction of a single, thin blast wave with clouds in the external
medium.  Large amplitude, short timescale variability occurs when the clouds have radii $r\ll
R/\Gamma$, where $R$ is the mean distance of a cloud from the GRB source and $\Gamma$ is
the blast-wave Lorentz factor.  Efficiencies $\gtrsim 10$\% require a large number of
small clouds, each with sufficiently large column densities to extract most of the
available blast-wave energy in the region of interaction. The number and duration of pulses
in the simulated GRB light curves are compared with the respective properties found in
 GRB light curves.  If GRB sources are surrounded by clouds with such properties, then short
timescale variability of GRBs can be obtained in the external shock model.
\end{abstract}


\section{Introduction}

The cosmological origin of GRBs has been established as a result of optical follow-up
observations of fading X-ray counterparts to GRBs discovered with the {\it Beppo-Sax}
mission (e.g., Costa et al.\ \markcite{cea97}1997;
van Paradijs et al.\ \markcite{vea97}1997;  Djorgovski et al.\ \markcite{dea97}1997).
The total energy released in GRB 970508 and GRB 971214 exceeds $10^{52}$ (e.g., Waxman
\markcite{waxman97}1997) and $10^{53}$ (Kulkarni et al.\ \markcite{kea98}1998) ergs,
respectively, if the emission is unbeamed. The time profiles of the X-ray afterglow
light curves are generally well fit by power laws (e.g.,
Piro et al.\ \markcite{pea98}1998; Feroci et al.\ \markcite{fea98}1998).  

These observations provide support for the fireball/blast-wave model of GRBs (Rees \& 
M\'esz\'aros \markcite{rm92}1992). In this model, a large quantity of energy released
in a small volume produces a GRB when the baryon loading of the fireball yields a
blast wave with Lorentz factor $\Gamma\sim 100$-300. Such values of $\Gamma$ are
required to avoid significant attenuation of gamma-rays through pair production processes
(e.g., Baring \& Harding \markcite{bh97}1997). Fireballs with larger and smaller baryon
loading are difficult to detect because of design limitations of past telescopes (Dermer,
Chiang, \& B\"ottcher \markcite{dcb98}1998). 

The success of the relativistic blast-wave
model is due largely to the relative simplicity with which it explains the temporal
dependence and intensity of the long wavelength GRB afterglows (e.g., Wijers, Rees, \&
M\'esz\'aros \markcite{wrm97}1997; Waxman \markcite{waxman97a}1997; Vietri
\markcite{vietri97}1997; Chiang \& Dermer \markcite{cd98}1998). Power-law decays in the
afterglow light curves result from the deceleration of the blast wave as it becomes
energized by sweeping up material from the circumburst medium (CBM).  This is termed the
external shock model of GRBs, and was originally introduced to provide a mechanism for
converting the energy of the blast wave into radiation during the prompt gamma-ray
luminous phase of GRBs (M\'esz\'aros \& Rees \markcite{mr93}1993).  

Recent research (Fenimore, Madras, \& Nayakshin \markcite{fmn96}1996, Sari \& Piran
\markcite{sp97}1997; Fenimore et al. \markcite{fea98}1998) indicates that the external
shock model faces  difficulties in explaining short timescale variability (STV) in
GRB light curves and afterglows as a consequence of special relativistic effects
resulting from the curvature of the blast wave shell.  In this {\it Letter}, we
construct a computer model to examine requirements of an external shock model to produce
STV.  In Section 2, we outline the temporal spreading problem due to the curvature of a
blast wave. Our numerical model used to simulate this system is described in Section 3,
and results are presented in Section 4.  If STV in GRB light curves is due to an external
shock model, then the required characteristics of such a model are outlined in Section 5.

\section{Variability from Localized Blast-Wave Emission Regions}

The STV problem in the external shock model is illustrated in Figure 1.  A blast wave,
expanding with Lorentz factor $\Gamma$, sweeps up material from a density inhomogeneity, or
``cloud,'' located at a
distance $R$ from the explosion site. We assume for simplicity that the cloud is spherical
with radius $r$. Let $\theta$ represent the angle between the directions from the GRB
explosion site to an observer and to the center of the cloud.  The blast wave first
interacts with the cloud at point 1, and stops interacting when it reaches point 4. The
largest angular extent of the blast wave energized by the cloud is defined by the
locations of points 2 and 3. The observer will record photon arrival times from each of
the four points. The time delay between photons emitted from points 1 and 4 is 
$$\Delta t_r = {2r\over \beta c}\; (1-\beta\cos\theta)\;(1+z)\; \equiv {2r(1+z)\over
\beta c \Gamma{\cal D}}\cong t_{\rm dur}[{r\over R}(1+\theta^2\Gamma^2)]
\; ,\eqno(1)$$
where the Doppler factor $\D = [\G(1-\b\cos\theta)]^{-1}$, $\b = (1-1/\G^2)^{1/2}$, and
$z$ is the redshift.  The quantity $t_{\rm dur} \equiv R_{\rm max}(1+z)/(\G^2c)$
represents the characteristic  duration of emission observed from a blast wave which
radiates as it passes through a medium of maximum extent
$R_{\rm max}> R$.  The expression on the right side of eq.\ (1) holds in the limit
$\G \gg 1$, implying that $\theta \ll 1$ because most of the radiation is observed from a
cone of angular extent$\lesssim 1/\G$ due to the Doppler effect. For constant $\Gamma$,
the time delay between photons emitted from points 2 and 3 is 
$$\Delta t_a = {2R\over c}\; \sin\theta\sin\chi\;(1+z)\cong t_{\rm dur}[{r\over
R}(2\Gamma^2\theta)]\; ,\eqno(2)$$
where we assume that $r/R\ll 1$ and $\G\gg 1$ in the right-most expression of eq.\ (2).

The term $\Delta t_r$ is a radial spreading time scale related to the observed interval
over which the blast wave passes through the radial extent of the cloud, and the term
$\Delta t_a$ is an angular spreading time scale related to the angular extent of
the cloud (see Fenimore et al.\ \markcite{fmn96}1996; Sari \& Piran \markcite{sp97}1997).
The shortest temporal duration from the blast-wave/cloud interaction is defined by the
maximum of these two time scales.  Note, however, that the pulse width from an
interaction event could be longer then the kinematic minimum time scale derived here if
the blast wave does not radiate its swept-up energy rapidly, or if the blast-wave
shell has thickness $\ell \gtrsim r$. In the limit that the blast wave is approximated by
expanding collisionless particles, $\ell \sim R/(2\G^2)$ (M\'esz\'aros, Laguna, \& Rees
\markcite{mlr93}1993).  The shell would be much thinner, however, if treated as a radiative
hydrodynamical fluid that is compressed by the pressure of the external medium (Blandford \& McKee 
\markcite{bm76}1976), and we
assume that this is the case here.

As can be seen by comparing eqs.(1) and (2), $\Delta t_a$ is much longer than $\Delta
t_r$ except within a very narrow cone of extent $\theta \lesssim 1/(2\G^2)$. If the blast
wave is uncollimated, much of the observed radiation is produced from interactions near
$\theta \sim 1/\G$ where $\Delta t_a/\Delta t_r \sim \G$.  To produce STV with
$\Delta t_a < \delta t\ll t_{\rm dur}$ therefore requires very small clouds with
$r/(R/\G) \lesssim (\delta t/t_{\rm dur})$. GRB time profiles commonly show pulses with
$ \delta t/t_{\rm dur} \lesssim 0.01 - 0.1$; thus very small clouds which
individually have a small surface covering factor are required for STV. The average
covering factor is the ratio of the total area of the clouds to the area of the beaming
cone, and is given by
$\approx I_c \pi r^2/[\pi (R_{\rm max}/\G)^2] \equiv
I_c f^2$, where $I_c$ is the number of clouds in the beaming cone and $f\equiv r/(R_{\rm
max}/\G)$. The efficiency for
extracting energy from a blast wave is therefore roughly equal to 
$$\zeta = \min(N_c/N_{cd},1)\;I_c\;f^2\; .\eqno(3)$$
Here $N_c$ is the column density of a cloud, and $N_{cd}$ is the column density that a
cloud must have to extract a significant fraction of the available energy in the
portion of the blast wave which interacts with the cloud. Here we implicitly assume that 
a large fraction of the blast wave energy is converted to nonthermal electrons and 
then to radiation. If this is not the case, then the efficiency is reduced and the
energy requirements increased accordingly. If the explosion energy $E_0 =
10^{54} E_{54}$ ergs is emitted uniformly in all directions from the explosion site, then
strong deceleration occurs when the swept-up relativistic cloud mass $\Gamma m_c$ equals the baryonic mass in the region that interacts with the cloud, where $m_c = 4\pi N_{c}  m_p r^2/3$ is the cloud mass. This implies
$N_{cd}
\cong 3E_0/(16\pi\G^2 m_pc^2 R^2) = 4.4\times 10^{16} E_{54}/\Gamma_{300}^2 R_{17}^2 $
cm$^{-2}$, where $\G = 300\G_{300}$ and $R = 10^{17} R_{17}$ cm.

In order to  extract the blast wave's energy efficiently by interacting with
clouds, we see from eq.\ (3) that a large number of clouds with thick columns (i.e, $N_c
\gtrsim N_{cd}$) are required to offset their small size.  If, for example, $f=0.01$,
then $\gtrsim 1000$ clouds with thick columns within the Doppler beaming cone will give an
efficiency  $\zeta \gtrsim 0.1$.  It has been argued
that pulses from the individual clouds would overlap (Sari \& Piran \markcite{sp97}1997;
Sumner \& Fenimore \markcite{sf98}1998) and fail to produce light curves with
large amplitude variability when the efficiencies are large. We have constructed a model
to assess this claim quantitatively.

\section{Description of the Model}

We use Monte Carlo methods to simulate the interaction of a spherically
expanding blast wave with clouds in the CBM. The clouds are randomly distributed in angle,
azimuth, and location. The radial distribution can be generally chosen but, for purposes
of illustration, we consider a uniform distribution in the range $ R < R_{\rm max} =
10^{17}$ cm. The mass and radius distributions of clouds can also be generally chosen,
but here we consider for clarity the case where all clouds have equal radii.  In
the results shown here, we assume that the clouds have thick columns. This means that $N_c
\gtrsim 4\times 10^{18} $ cm$^{-2}$ for clouds with $R>10^{16}$ cm; closer clouds
require larger column densities, but the number of such clouds is extremely small.  We
employ a blocking factor to account for occulted clouds.  After the locations and radii
of the clouds are selected, the program determines which clouds have been completely
occulted and eliminates them. When partial occultation occurs, we weight the contribution
of that cloud based on the fraction of the area that is occulted.

We use the following expression to represent the angle-dependent flux density
$S$(ergs cm$^{-2}$ s$^{-1} \eo^{-1}$) measured at observer time $t_{\rm obs}$
from that portion of the curved blast wave
surface which is energized by its interaction with a cloud:
$$S(\eo,\to;\Omega_{\rm obs}) = K\;{\D^3(1+z) J(\e )\over 4\pi d_L^2}\; 
{4({\ln 2 \over \pi})^{1/2}}e^{-(t_c-t_{\rm obs})^{2}/
2\sigma^{2}}\;,\eqno(4)$$
where
$$ K = \cases{ ~~~~~~~~~~~1 ~~~~~~~~~~~~~~~~~~~~~~~~~~~~~~~~~{\rm and}~~ \Delta t =
\Delta t_r ,& if $\theta \leq 1/(2\G^2)$;
\cr\cr 
  r(R\Gamma\D\sin\theta\sin\chi )^{-1}\; \cong \; (\G\theta\D)^{-1}~~{\rm and}
~~\Delta t =
\Delta t_a , & if $\theta>1/(2\G^2)$ \cr}\eqno(5)$$
(see Dermer, Sturner, \& Schlickeiser \markcite{dss97}1997).  In eq.\ (4), We approximate
the pulse duration by a Gaussian function with $\sigma=\Delta
t / (4\sqrt {2 \ln 2})$, which gives a FWHM pulse duration =$\Delta t /2$. In eq.\ (5),
$\eo=h\nu_{\rm obs}/m_ec^2 = \D
\e/(1+z)$ is the observed dimensionless photon energy and $\e $ is the photon energy
measured in the comoving frame of the blast wave, $t_c = (1+z) R(1-\b
\cos\theta)/(\b c)$ when $\Delta t_r > \Delta t_a$, and $t_c = (1+z) R(1-\b
\cos\theta\cos\chi)/(\b c)$ when $\Delta t_r <\Delta t_a$ The quantity
$J(\e )$ represents the spectral power in the comoving frame. The above result is obtained 
by noting that in the limit
$r\ll R/\Gamma$, the beaming factor for the time-integrated fluence from the
radiating portion of the blast wave energized by the interaction is the same as that for a
spherical ball of plasma which radiates for a comoving period of time equal to 
$t_2 - t_1 = 2r/\beta\G c$.

We approximate the comoving radiant emission by a power law such that $J(\e) =
J_0\e^{-\alpha}$ in the energy range $\e_1 \leq \e \leq \e_2$, where $\alpha$ is the
photon energy index.  To set the normalization for $J_0$, we note that due to the
increasing surface area of the expanding blast wave, the maximum energy that could be
radiated as a result of the blast wave/cloud interaction is just $E_c \cong [E_0
r^2/(4R^2)]\min(N_c/N_{cd},1)$.  We assume that this energy is radiated promptly, that
is, the interaction takes place in the radiative regime. The relationship between the
measured energy fluence $F_E$ (ergs cm$^{-2}$) and the total energy emitted is $F_E =
(1+z)E_c/(4\pi d_L^2)$, where $d_L$ is the luminosity distance.  From this relation, we
obtain the result $J_0 = f_\alpha[E_c/ \G (t_2- t_1)]\;=  f_\alpha (c
E_c/2r)$, where $f_\alpha \equiv (1-\alpha) / (\e_2^{1-\alpha} -
\e_1^{1-\alpha})$.

\section{Results}

In our calculations, we choose parameter values motivated by the observations of the
candidate host galaxy of GRB 971214 at redshift $z = 3.42$ (Kulkarni et al.\
\markcite{kea98}1998).  Its total energy release, including prompt radiation
and afterglow, could exceed $10^{54}$ ergs if the burst source is unbeamed.
Thus we choose a value of
$\partial E/\partial \Omega = 10^{53}$ ergs sr$^{-1}$ for our standard GRB energy release,
which implies a value of $E_{54} = 1.3$ if the burst energy is isotropically radiated.
For the calculations shown here, we let $\Gamma_{300} = 1$,
$z= 1$ (which implies $d_L = 1.67\times 10^{28}$ cm for a Hubble constant of 65 km s$^{-1}$ Mpc$^{-1}$ and $q_0 = 1/2$),
$\alpha = 1$, $\e_1 = 10^{-4}$, and $\e_2 = 10^2$.  In order to conserve computing time,
all clouds are placed at angles $\theta < k_\theta/\Gamma$ with $k_\theta = 3$, as
we find that clouds located at larger angles make a negligible contribution to the
observed flux due to the strong Doppler beaming. The clouds are 
located in a shell between $R_{\rm min} = 10^{16}$ and $R_{\rm max} = 10^{17}$ cm.

Fig.\ 2 shows results of our calculations for a range of different cloud sizes at $\eo = 1$. 
The number of clouds within the Doppler beaming cone $\theta < 1/\Gamma$ is chosen in order
to obtain a 10\% efficiency as defined by eq.\ (3).  In curves (a), (b), and (c) we let $f =
r/(R_{\rm max}/\Gamma) = 0.1, 0.03$, and 0.01, implying a total number of clouds
$k_\theta^2 I_c = 9 I_c = $ 90, 1000, and 9000 in the respective calculations. From the
expression for
$N_{cd}$ following eq.\ (3), this implies cloud densities $n_c \gtrsim 100 E_{54}/(\G_{300}
R_{17}^3 f)$ cm$^{-3}$.  By examining the curves, one can verify that the calculated energy
fluences agree with the expected energy fluence
$\cong 0.1 (1+z) 4\pi\times 10^{53}$ ergs sr$^{-1}/[4\pi d_L^2
\ln(\e_2/\e_1)] \sim 5\times 10^{-6}$ ergs cm$^{-2}$ for a 10\% efficiency.

From eq.\ (2), we see that a typical pulse duration is $\approx
\Delta t_a \approx 2 r(1+z)/(\G c)$ when $\theta \sim 1/\G$. The cloud size in curve (a)
has a value of $r = 3.3\times 10^{13}$ cm, implying
$\delta t = 15$ s when $\theta \sim 1/\G$, comparable to the value of $t_{\rm dur} = 74$
s.  Thus only limited structure in the light curves are seen.  For the cases with smaller
clouds shown in curves (b) and (c), the light curves become increasingly variable. 
In many but not all cases, a very narrow pulse with large amplitude is obtained as a result of a
cloud which happens to be located within an angle $\theta \lesssim 1/\Gamma^2$ of the
observer's line-of-sight. These early bright peaks might not be so apparent if the
blast-wave thickness also determines the pulse duration. The light curves are most variable during the period
 $\approx (1+z)( R_{\rm max}-R_{\rm min})/(2\G^2 c)\sim 33$ s.  The
tendency for longer pulses to be found in the simulated
light curves at later times is an artifact of the bounded cloud ranges in the numerical
simulation. This effect is ameliorated when more complicated cloud profiles without sharp
boundaries are considered. All these  simulated GRBs would be detectable with BATSE
on {\it CGRO} out to redshifts $z\approx 4$, noting that BATSE triggers on a $\nu F_\nu$
flux of $\gtrsim 10^{-7}$ ergs cm$^{-2}$ s$^{-1}$. Higher efficiency events could be
detected from sources at even larger redshifts (see inset to Fig. 2).

The mean number and FWHM durations of pulses for an ensemble of GRBs corresponding to cases 
(a), (b) and (c) in Fig. 2 are 7.9, 19 and 42, and 4.8 s, 1.5 s and 0.42 s, respectively.
We consider only those pulses which either rose from or decayed beyond 50\% of
their peak value, and used twice the half-maximum duration when only one of the aforementioned cases was
satisfied.  Norris et al.\ (\markcite{nea96}1996) has decomposed 41 bright GRB light curves to 
find a distribution of pulse durations in the range 0.2-2 s. As shown in the right inset to Fig.
2, cloud sizes intermediate to cases (b) and (c) are in accord with the distribution deduced
by Norris et al. A distribution of cloud sizes and the addition of background noise in the
simulations could improve the comparison.  

\section{Discussion and Summary}

In the external shock model considered here, STV results from the energy of a blast wave
being efficiently extracted by sweeping up matter from the CBM in the form of
small clouds (compare the approach of Shaviv
\& Dar \markcite{sd95}1995).  If the clouds are distributed uniformly around the burst
source, then a total cloud mass of
$$M_c \cong 4\Gamma^2 I_c \cdot {4\pi\over 3} m_p N_{c, \min} r^2 \gtrsim 3\times 10^{-4}
\zeta R_{17}^2\;({E_{54}\over \G_{300}^2 R_{16}^2})\;M_\odot \eqno(6)$$
is required.  Here the minimum cloud column density $N_{c,\min}$ is derived for clouds
at $10^{16}R_{16}$ cm. In contrast, the minimum mass of the surrounding intercloud medium
(ICM) required to extract a significant fraction of the blast
wave's energy is $E_0/(\G^2 c^2) \approx 6\times 10^{-6} E_{54}/\G^2_{300}~M_\odot $.  More
mass must be in the form of small clouds to ensure that their columns are thick at all values
of $R$, but the total mass is remarkably small in  either case. The contrast between the
ICM density $n_{ICM}$ and the cloud density $n_c$ required to obtain STV is $n_{ICM}/n_c
\ll (4f\zeta/\Gamma) \ll 1 $. Whether such clouds can be stably confined or must be
continuously reformed is unclear.

We consider arguments presented against the external shock model (Fenimore, Ramirez, \&
 Sumner \markcite{frs98}1998; Fenimore et al. \markcite{fmn96}1996; Sari \& Piran
\markcite{sp97}1997; Sumner \& Fenimore \markcite{sf98}1998). The strongest objection is
that the overlapping light curves will ``wash out" large amplitude variabilty.  According
to eq.\ (2), most clouds produce pulses of duration $\delta t \gtrsim 2ft_{\rm
dur}(\G\theta)$. Because most clouds in the Doppler cone are located at $\theta \sim
1/\G$,  STV only occurs when $\delta t/ t_{\rm dur} \sim f \ll 1$. To obtain a given
efficiency, eq.\ (3) shows that $I_c = \zeta/f^2$  clouds with thick columns are
required. The number of overlapping clouds in the time element $\delta t$ is $\approx I_c
(\delta t/t_{\rm dur}) \approx f I_c \approx
\zeta/f$. This would apparently yield statistical fluctuations of order
$(f/\zeta)^{1/2}$ which, for
$f = 0.01$ and $\zeta = 0.1$, is at the 30\% level rather than at the factor-of-2 level
often seen in GRB light curves.

This argument overlooks the fact that the small fraction of clouds at $\theta \ll
1/\Gamma$ makes a disproportionate contribution to the amplitude and variability of GRB
light curves. Only 1\% of all clouds in the Doppler beaming cone lie in the range $0<
\theta < 1/(10\G)$, yet the duration of pulses from these clouds is shorter by an
order-of-magnitude than the duration of pulses from clouds at $\theta = 1/\G$. 
Moreover, the amplitude of the pulses varies according
to the factor $\sim \D^{(2+\alpha)}/(\G\theta)$, which represents a factor $\approx 80$
for clouds at $\theta < 1/(10\G)$ when compared to clouds at $\theta \cong 1/\G$. We
therefore see that 1\% of the clouds produce $\approx 8$\% of the total fluence in
very short duration, very large amplitude pulses. The amplitude enhancements continue to
increase $\propto 1/\theta$ until $\theta
\lesssim 1/(2\G^2)$, or until the pulse width becomes limited by non-kinematic effects.
If clouds are found in layers, then the enhanced contribution of on-axis clouds might also overcome the difficulty in producing gaps in GRB light curves (Fenimore, Ramirez, \&
Sumner \markcite{frs98}1998).

The efficiency $\zeta$ defined in eq.\ (3) represents the most optimistic case where
all the blast wave energy in the region energized by the cloud is transformed into
radiation. Because the blast wave decelerates strongly in the region of interaction, STV
can be produced through Doppler deboosting even if the energized electrons do not lose most
of their energy through synchrotron processes (Chiang \markcite{jc99a,99b}1999a,b). In this
case, of course, the efficiency is much less than obtained through eq.\ (3).

In summary, we have examined a model in accord with the third scenario described by Fenimore
et al.\ \markcite{fea96}(1996),
 where STV in GRB light curves is produced by inhomogeneties in the CBM. This breaks the
condition of local spherical symmetry, shown by Fenimore et al.\ to produce light
curves inconsistent with observations, but requires the presence of many small clouds with
thick
 columns. As shown here, STV in GRB light curves can be produced by the external shock model
under
 such conditions. Further comparisons of GRB data with model light curves will be necessary,
however, to establish whether GRB light curves are tomographic images of the density
distributions of the medium surrounding the sources of GRBs

\acknowledgements We thank M. B\"ottcher, J. Chiang, E. Fenimore, R. Sari and the anonymous
referee for comments and discussions.  The work of CDD was supported by the Office of Naval
Research and the {\it Compton Gamma Ray Observatory} Guest Investigator Program. The work of
KEM was made possible through the Science and Engineering Apprentice Program of George
Washington University.

\eject

\begin{figure}
\epsfxsize=\hsize
\epsfbox{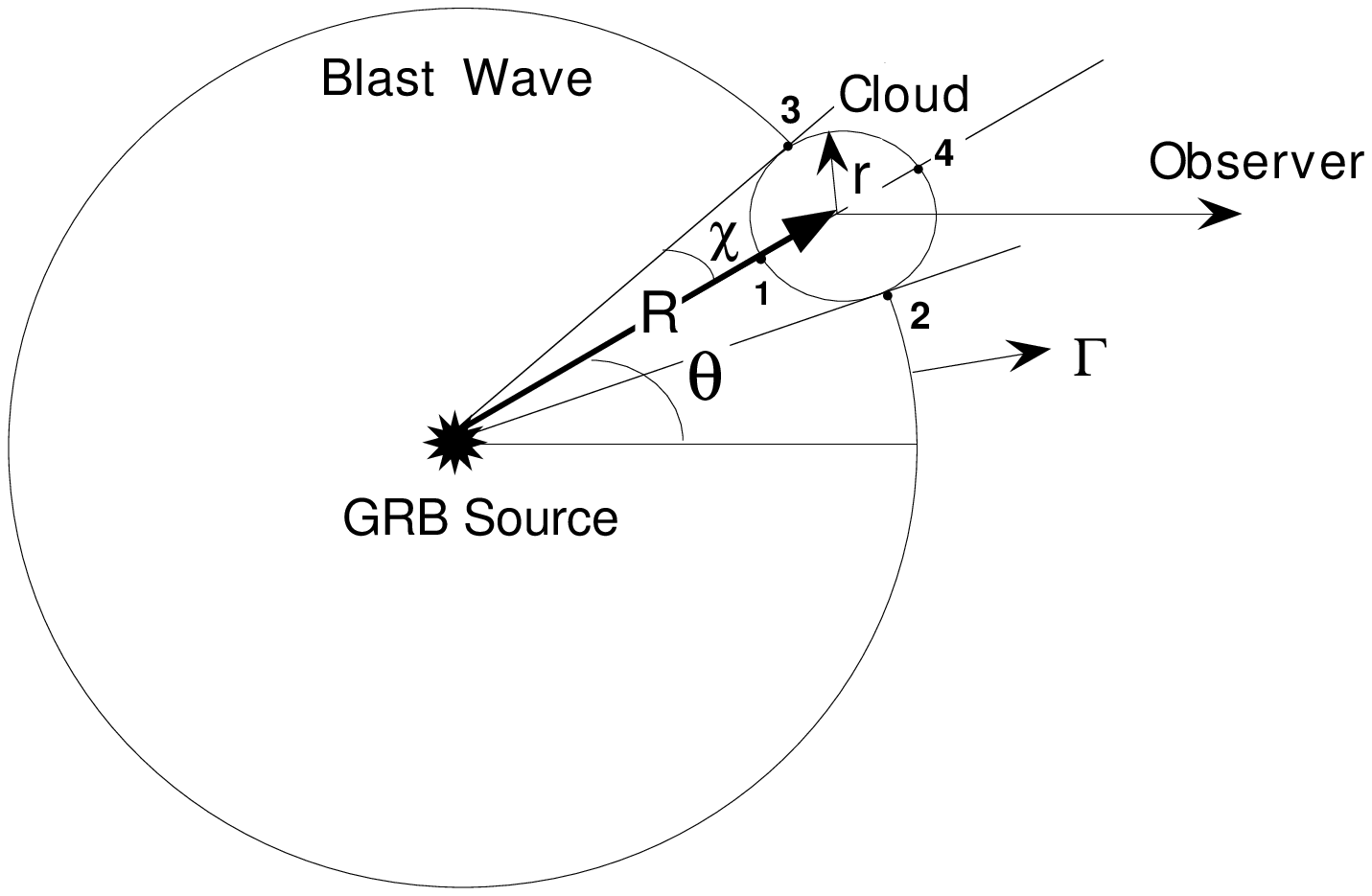}
\caption[]{Geometry of the interaction between a blast
wave and a cloud in the circumburst medium. }
\end{figure}

\begin{figure}
\epsfbox{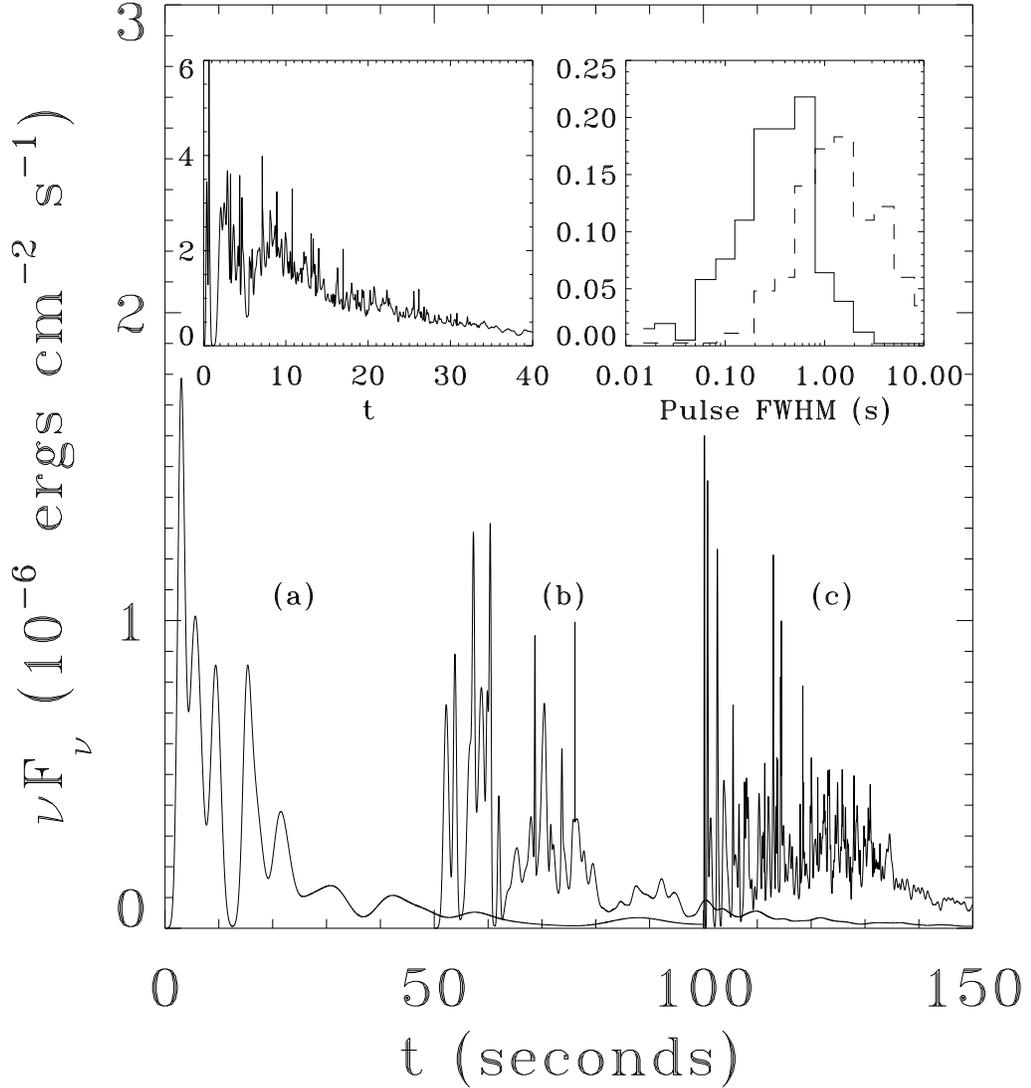}
\caption[]{Simulated light curves from blast-wave/cloud interactions. The cloud radii
range from $3.3\times 10^{13}$ cm in curve (a) to $3.3\times 10^{12}$ cm in curve
(c), and the total number of clouds is estimated from eq.\ (3) in order to extract 10\%
of the blast wave energy (see text).   Successive curves are offset by 50 s
each for clarity.  Left inset shows a calculation for 
$3.3\times 10^{12}$ cm clouds  giving $\sim 100$\% efficiency. Right inset shows the
distributions of pulse widths, normalized to unity, for curves (case b; dashed histogram) and (case c;
solid histogram). The early narrow peak in graph (c) extends to
$3\times 10^{-6}$ ergs cm$^{-2}$ s$^{-1}$, but is covered by the inset. }
\end{figure}
\end{document}